\documentclass[11pt]{article}
\usepackage{graphicx}

\begin{document}

\def\tfrac{\textstyle\frac}
\def\beq{\begin{equation}}
\def\eeq{\end{equation}}

\begin{center}
{\LARGE Spacetime and Euclidean Geometry}\footnote{This article 
is dedicated to Michael P. Ryan on the occasion of his sixtieth birthday. 
Mike's passion for, and deft practice of, both geometry and 
pedagogy is legendary at Maryland. 
We are pleased with this opportunity to present
our pedagogical effort to elucidate the geometry of Minkowski spacetime, 
the most homogeneous of cosmologies.}\\
\vspace{5mm}
{\large Dieter Brill$^*$ and Ted Jacobson$^{*,\dagger}$}\\
\vspace{5mm}
{\it \small $^*$University of Maryland, College Park, MD 20742-4111}\\
{\it \small $^\dagger$Institut d'Astrophysique de Paris, 98bis bvd. Arago, 
75014 Paris, France}
\end{center}

\begin{abstract}
Using only the principle of relativity and Euclidean geometry we
show in this pedagogical article that the square of proper time or
length in a two-dimensional spacetime diagram is proportional to
the Euclidean area of the corresponding causal domain. We use this
relation to derive the Minkowski line element by two geometric
proofs of the {\it spacetime Pythagoras theorem.}
\end{abstract}

\subsection*{Introduction}
\label{Intro} Spacetime diagrams are helpful for understanding
relativity since they focus attention on the invariant relations
between events, light rays, observers, etc. rather than on
coordinate dependent quantities. An inherent limitation of such
diagrams  is that, in general, the Euclidean lengths of lines in
the diagram do not correspond to proper time or proper length in
spacetime. In this pedagogical article we use the principle of
relativity, together with Euclidean geometry, to show that
nevertheless the {\it square} of proper time or length of 
a line segment is proportional to the Euclidean {\it area} of the
corresponding causal domain. This observation allows visual
interpretation of relativistic effects, such as time dilation and
the twin effect. We use this relation between Minkowski interval
and Euclidean area to derive the Minkowski line element by proving
the {\it spacetime Pythagoras theorem}.

\subsection*{Minkowski space and Euclidean space}
\label{MinkEuc}

\begin{figure}
\unitlength 1mm 
\linethickness{0.4pt}
\ifx\plotpoint\undefined\newsavebox{\plotpoint}\fi 
\begin{picture}(59,50)(-20,0)
\put(40,30){\circle*{1}} \put(37,29){\makebox(0,0)[cc]{$p$}}
\put(40,30){\line(3,-1){18}}
\put(58,24){\circle*{1}}
\put(54,19){\makebox(0,0)[lc]{spacelike}}
\put(53,20){\vector(-1,3){2}}
\put(51,48){\circle*{1}}
\put(45,49){\makebox(0,0)[rc]{timelike}}
\put(46,48){\vector(1,-3){1.67}}
\put(57,40){\makebox(0,0)[lc]{lightlike}}
\put(56,40){\vector(-1,0){5}}
\put(20,50){\line(0,1){0}}
\put(40,30){\line(3,5){10.8}}
\multiput(19.93,9.93)(.033051,.033051){20}{\line(1,0){.033051}}
\multiput(21.25,11.25)(.033051,.033051){20}{\line(0,1){.033051}}
\multiput(22.57,12.57)(.033051,.033051){20}{\line(1,0){.033051}}
\multiput(23.9,13.9)(.033051,.033051){20}{\line(0,1){.033051}}
\multiput(25.22,15.22)(.033051,.033051){20}{\line(1,0){.033051}}
\multiput(26.54,16.54)(.033051,.033051){20}{\line(0,1){.033051}}
\multiput(27.86,17.86)(.033051,.033051){20}{\line(0,1){.033051}}
\multiput(29.18,19.18)(.033051,.033051){20}{\line(0,1){.033051}}
\multiput(30.51,20.51)(.033051,.033051){20}{\line(1,0){.033051}}
\multiput(31.83,21.83)(.033051,.033051){20}{\line(0,1){.033051}}
\multiput(33.15,23.15)(.033051,.033051){20}{\line(0,1){.033051}}
\multiput(34.47,24.47)(.033051,.033051){20}{\line(0,1){.033051}}
\multiput(35.79,25.79)(.033051,.033051){20}{\line(1,0){.033051}}
\multiput(37.12,27.12)(.033051,.033051){20}{\line(0,1){.033051}}
\multiput(38.44,28.44)(.033051,.033051){20}{\line(0,1){.033051}}
\multiput(39.76,29.76)(.033051,.033051){20}{\line(1,0){.033051}}
\multiput(41.08,31.08)(.033051,.033051){20}{\line(0,1){.033051}}
\multiput(42.4,32.4)(.033051,.033051){20}{\line(0,1){.033051}}
\multiput(43.73,33.73)(.033051,.033051){20}{\line(0,1){.033051}}
\multiput(45.05,35.05)(.033051,.033051){20}{\line(0,1){.033051}}
\multiput(46.37,36.37)(.033051,.033051){20}{\line(1,0){.033051}}
\multiput(47.69,37.69)(.033051,.033051){20}{\line(0,1){.033051}}
\multiput(49.01,39.01)(.033051,.033051){20}{\line(0,1){.033051}}
\multiput(50.34,40.34)(.033051,.033051){20}{\line(1,0){.033051}}
\multiput(51.66,41.66)(.033051,.033051){20}{\line(0,1){.033051}}
\multiput(52.98,42.98)(.033051,.033051){20}{\line(0,1){.033051}}
\multiput(54.3,44.3)(.033051,.033051){20}{\line(0,1){.033051}}
\multiput(55.62,45.62)(.033051,.033051){20}{\line(0,1){.033051}}
\multiput(56.95,46.95)(.033051,.033051){20}{\line(0,1){.033051}}
\multiput(58.27,48.27)(.033051,.033051){20}{\line(0,1){.033051}}
\multiput(58.93,10.93)(-.033051,.033051){20}{\line(0,1){.033051}}
\multiput(57.61,12.25)(-.033051,.033051){20}{\line(0,1){.033051}}
\multiput(56.29,13.57)(-.033051,.033051){20}{\line(0,1){.033051}}
\multiput(54.96,14.9)(-.033051,.033051){20}{\line(-1,0){.033051}}
\multiput(53.64,16.22)(-.033051,.033051){20}{\line(0,1){.033051}}
\multiput(52.32,17.54)(-.033051,.033051){20}{\line(0,1){.033051}}
\multiput(51,18.86)(-.033051,.033051){20}{\line(0,1){.033051}}
\multiput(49.68,20.18)(-.033051,.033051){20}{\line(-1,0){.033051}}
\multiput(48.35,21.51)(-.033051,.033051){20}{\line(0,1){.033051}}
\multiput(47.03,22.83)(-.033051,.033051){20}{\line(0,1){.033051}}
\multiput(45.71,24.15)(-.033051,.033051){20}{\line(0,1){.033051}}
\multiput(44.39,25.47)(-.033051,.033051){20}{\line(0,1){.033051}}
\multiput(43.07,26.79)(-.033051,.033051){20}{\line(-1,0){.033051}}
\multiput(41.74,28.12)(-.033051,.033051){20}{\line(0,1){.033051}}
\multiput(40.42,29.44)(-.033051,.033051){20}{\line(0,1){.033051}}
\multiput(39.1,30.76)(-.033051,.033051){20}{\line(-1,0){.033051}}
\multiput(37.78,32.08)(-.033051,.033051){20}{\line(0,1){.033051}}
\multiput(36.46,33.4)(-.033051,.033051){20}{\line(0,1){.033051}}
\multiput(35.13,34.73)(-.033051,.033051){20}{\line(0,1){.033051}}
\multiput(33.81,36.05)(-.033051,.033051){20}{\line(0,1){.033051}}
\multiput(32.49,37.37)(-.033051,.033051){20}{\line(-1,0){.033051}}
\multiput(31.17,38.69)(-.033051,.033051){20}{\line(0,1){.033051}}
\multiput(29.84,40.01)(-.033051,.033051){20}{\line(0,1){.033051}}
\multiput(28.52,41.34)(-.033051,.033051){20}{\line(-1,0){.033051}}
\multiput(27.2,42.66)(-.033051,.033051){20}{\line(0,1){.033051}}
\multiput(25.88,43.98)(-.033051,.033051){20}{\line(0,1){.033051}}
\multiput(24.56,45.3)(-.033051,.033051){20}{\line(0,1){.033051}}
\multiput(23.23,46.62)(-.033051,.033051){20}{\line(0,1){.033051}}
\multiput(21.91,47.95)(-.033051,.033051){20}{\line(0,1){.033051}}
\multiput(20.59,49.27)(-.033051,.033051){20}{\line(0,1){.033051}}
\put(58.93,48.93){\line(0,1){0}}
\end{picture}
\vskip-5\unitlength \caption{The different directions in Minkowski
space.}
\label{spacetime}
\end{figure}
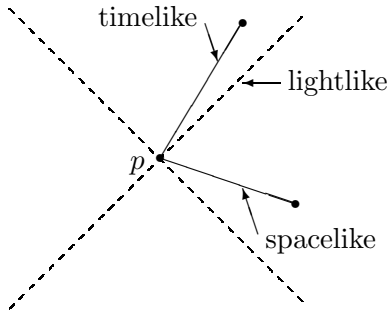

In a two-dimensional spacetime diagram a spacetime, i.e. a
Minkowski space, is represented on a Euclidean plane. This is
possible since, like the points in the Euclidean plane, the events
in spacetime can be labeled by pairs of real numbers, for example
time and space coordinates. What other properties of Minkowski
space can this mapping faithfully reproduce?

Euclidean and Minkowski spaces admit identical translation
symmetry groups that act in the same way on their respective
spaces. Therefore the mapping can be chosen to preserve the
translation symmetry. A straight line can be characterized as a
curve that is sent to itself by the translations in one direction.
Since the translations are preserved by the mapping, so are the
straight lines.\footnote{For a more systematic treatment of the consequences of translation
symmetry, as well as an axiomatic development of all of Minkowski Space
geometry, see Ref.~ \cite{WilsonLewis}. For a different axiomatization,
based on the relation of causal connection between points, 
see Ref.~\cite{Robb}.}

Given the 
Minkowski interval, i.e.~the proper time or length 
between two points on a line, the interval
between any two other points on the same line is determined by the
translational symmetry. The Euclidean distance along a line in a
spacetime diagram is therefore {\it proportional} to the
corresponding 
Minkowski interval.
The obstacle to
representing all aspects of Minkowski geometry in a Euclidean
diagram is that the proportionality factor depends upon the line.

This obstacle arises because, unlike in Euclidean space, not all
Minkowski lines are equivalent (Fig.~\ref{spacetime}). That is,
Minkowski space is not {\it isotropic}. According to relativity,
the lightrays through a point $p$ are determined independent of
the motion of any source. They therefore divide up the spacetime
into four intrinsic regions: the future, past, right space, and
left space of $p$. A {\it timelike} line segment has one endpoint
to the future of the other, and represents the inertial motion of
a free particle. For a {\it spacelike} segment neither endpoint
lies to the future of the other. The borderline case, 
with endpoints connected by a light ray, is
called a {\it lightlike} or {\it null} segment.
The light rays on all diagrams in this paper are shown as dashed
lines, while the timelike and spacelike lines are solid.

In drawing a spacetime diagram one must select two independent
directions to represent the light rays. This choice breaks the
Euclidean rotation symmetry. It is common to orient the diagram so
that the timelike line bisecting the angle between the light rays
runs vertically up the diagram. The vertical direction then
corresponds to pure time translations in some particular frame,
while the horizontal direction corresponds to pure space
translations in that frame. It is also common to choose the
relation between the vertical and horizontal  scales so that the
two sets of light rays are perpendicular to each other in the
Euclidean sense, and therefore make angles of 45 degrees from the
vertical. With such choices of scaling, horizontal and vertical
segments of the the same Euclidean length represent intervals of
Minkowski length and time with ratio equal to the speed of light.

Although common, the choice of a right angle between light rays is not
mandatory. The constructions and proofs in this paper could all be
carried out with an arbitrary angle, but we shall adopt the right angle
because it makes the diagrams and proofs a little easier to follow,
and because it is the standard and familiar choice.

\subsection*{Squares and triangles}

In this section we introduce some concepts basic to space-time
geometry that will be used in the following.

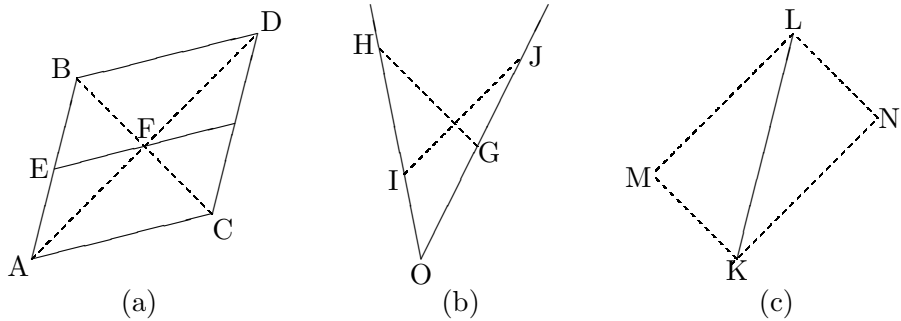
\begin{figure}[h]
\unitlength 0.75mm \linethickness{0.4pt}
\ifx\plotpoint\undefined\newsavebox{\plotpoint}\fi 
\begin{picture}(160,55)(15,0)
\put(92,10){\line(-1,5){9}} \put(92,10){\line(1,2){22.5}}
\put(92,9){\makebox(0,0)[ct]{O}}
\put(102,29){\makebox(0,0)[lc]{G}}
\put(83.75,48){\makebox(0,0)[rc]{H}}
\put(88,24){\makebox(0,0)[rc]{I}}
\put(111,46){\makebox(0,0)[lc]{J}}
\put(148,10){\line(1,4){10}}
\multiput(101.93,29.93)(-.032143,.032143){20}{\line(0,1){.032143}}
\multiput(100.64,31.22)(-.032143,.032143){20}{\line(0,1){.032143}}
\multiput(99.36,32.5)(-.032143,.032143){20}{\line(0,1){.032143}}
\multiput(98.07,33.79)(-.032143,.032143){20}{\line(0,1){.032143}}
\multiput(96.79,35.07)(-.032143,.032143){20}{\line(0,1){.032143}}
\multiput(95.5,36.36)(-.032143,.032143){20}{\line(0,1){.032143}}
\multiput(94.22,37.64)(-.032143,.032143){20}{\line(-1,0){.032143}}
\multiput(92.93,38.93)(-.032143,.032143){20}{\line(0,1){.032143}}
\multiput(91.64,40.22)(-.032143,.032143){20}{\line(-1,0){.032143}}
\multiput(90.36,41.5)(-.032143,.032143){20}{\line(0,1){.032143}}
\multiput(89.07,42.79)(-.032143,.032143){20}{\line(0,1){.032143}}
\multiput(87.79,44.07)(-.032143,.032143){20}{\line(0,1){.032143}}
\multiput(86.5,45.36)(-.032143,.032143){20}{\line(0,1){.032143}}
\multiput(85.22,46.64)(-.032143,.032143){20}{\line(-1,0){.032143}}
\multiput(88.93,24.93)(.032813,.032813){20}{\line(0,1){.032813}}
\multiput(90.24,26.24)(.032813,.032813){20}{\line(0,1){.032813}}
\multiput(91.55,27.55)(.032813,.032813){20}{\line(0,1){.032813}}
\multiput(92.87,28.87)(.032813,.032813){20}{\line(0,1){.032813}}
\multiput(94.18,30.18)(.032813,.032813){20}{\line(0,1){.032813}}
\multiput(95.49,31.49)(.032813,.032813){20}{\line(0,1){.032813}}
\multiput(96.8,32.8)(.032813,.032813){20}{\line(0,1){.032813}}
\multiput(98.12,34.12)(.032813,.032813){20}{\line(0,1){.032813}}
\multiput(99.43,35.43)(.032813,.032813){20}{\line(0,1){.032813}}
\multiput(100.74,36.74)(.032813,.032813){20}{\line(0,1){.032813}}
\multiput(102.05,38.05)(.032813,.032813){20}{\line(0,1){.032813}}
\multiput(103.37,39.37)(.032813,.032813){20}{\line(0,1){.032813}}
\multiput(104.68,40.68)(.032813,.032813){20}{\line(0,1){.032813}}
\multiput(105.99,41.99)(.032813,.032813){20}{\line(0,1){.032813}}
\multiput(107.3,43.3)(.032813,.032813){20}{\line(0,1){.032813}}
\multiput(108.62,44.62)(.032813,.032813){20}{\line(0,1){.032813}}
\put(99,2){\makebox(0,0)[cc]{(b)}}
\put(155,2){\makebox(0,0)[cc]{(c)}}
\put(148,8){\makebox(0,0)[cc]{K}}
\put(130.5,24.5){\makebox(0,0)[cc]{M}}
\put(175,35){\makebox(0,0)[cc]{N}}
\put(158,52.5){\makebox(0,0)[cc]{L}}
\multiput(147.93,9.93)(-.032895,.032895){19}{\line(0,1){.032895}}
\multiput(146.68,11.18)(-.032895,.032895){19}{\line(0,1){.032895}}
\multiput(145.43,12.43)(-.032895,.032895){19}{\line(0,1){.032895}}
\multiput(144.18,13.68)(-.032895,.032895){19}{\line(0,1){.032895}}
\multiput(142.93,14.93)(-.032895,.032895){19}{\line(0,1){.032895}}
\multiput(141.68,16.18)(-.032895,.032895){19}{\line(0,1){.032895}}
\multiput(140.43,17.43)(-.032895,.032895){19}{\line(0,1){.032895}}
\multiput(139.18,18.68)(-.032895,.032895){19}{\line(0,1){.032895}}
\multiput(137.93,19.93)(-.032895,.032895){19}{\line(0,1){.032895}}
\multiput(136.68,21.18)(-.032895,.032895){19}{\line(0,1){.032895}}
\multiput(135.43,22.43)(-.032895,.032895){19}{\line(0,1){.032895}}
\multiput(134.18,23.68)(-.032895,.032895){19}{\line(0,1){.032895}}
\multiput(146.93,8.93)(.0325,.0325){20}{\line(1,0){.0325}}
\multiput(148.23,10.23)(.0325,.0325){20}{\line(0,1){.0325}}
\multiput(149.53,11.53)(.0325,.0325){20}{\line(1,0){.0325}}
\multiput(150.83,12.83)(.0325,.0325){20}{\line(1,0){.0325}}
\multiput(152.13,14.13)(.0325,.0325){20}{\line(1,0){.0325}}
\multiput(153.43,15.43)(.0325,.0325){20}{\line(1,0){.0325}}
\multiput(154.73,16.73)(.0325,.0325){20}{\line(0,1){.0325}}
\multiput(156.03,18.03)(.0325,.0325){20}{\line(1,0){.0325}}
\multiput(157.33,19.33)(.0325,.0325){20}{\line(1,0){.0325}}
\multiput(158.63,20.63)(.0325,.0325){20}{\line(1,0){.0325}}
\multiput(159.93,21.93)(.0325,.0325){20}{\line(1,0){.0325}}
\multiput(161.23,23.23)(.0325,.0325){20}{\line(0,1){.0325}}
\multiput(162.53,24.53)(.0325,.0325){20}{\line(1,0){.0325}}
\multiput(163.83,25.83)(.0325,.0325){20}{\line(1,0){.0325}}
\multiput(165.13,27.13)(.0325,.0325){20}{\line(0,1){.0325}}
\multiput(166.43,28.43)(.0325,.0325){20}{\line(1,0){.0325}}
\multiput(167.73,29.73)(.0325,.0325){20}{\line(0,1){.0325}}
\multiput(169.03,31.03)(.0325,.0325){20}{\line(1,0){.0325}}
\multiput(170.33,32.33)(.0325,.0325){20}{\line(1,0){.0325}}
\multiput(171.63,33.63)(.0325,.0325){20}{\line(0,1){.0325}}
\multiput(172.93,34.93)(-.032895,.032895){19}{\line(0,1){.032895}}
\multiput(171.68,36.18)(-.032895,.032895){19}{\line(0,1){.032895}}
\multiput(170.43,37.43)(-.032895,.032895){19}{\line(0,1){.032895}}
\multiput(169.18,38.68)(-.032895,.032895){19}{\line(0,1){.032895}}
\multiput(167.93,39.93)(-.032895,.032895){19}{\line(0,1){.032895}}
\multiput(166.68,41.18)(-.032895,.032895){19}{\line(0,1){.032895}}
\multiput(165.43,42.43)(-.032895,.032895){19}{\line(0,1){.032895}}
\multiput(164.18,43.68)(-.032895,.032895){19}{\line(0,1){.032895}}
\multiput(162.93,44.93)(-.032895,.032895){19}{\line(0,1){.032895}}
\multiput(161.68,46.18)(-.032895,.032895){19}{\line(0,1){.032895}}
\multiput(160.43,47.43)(-.032895,.032895){19}{\line(0,1){.032895}}
\multiput(159.18,48.68)(-.032895,.032895){19}{\line(0,1){.032895}}
\multiput(157.93,49.93)(-.032895,-.032895){20}{\line(-1,0){.032895}}
\multiput(156.61,48.61)(-.032895,-.032895){20}{\line(-1,0){.032895}}
\multiput(155.3,47.3)(-.032895,-.032895){20}{\line(0,-1){.032895}}
\multiput(153.98,45.98)(-.032895,-.032895){20}{\line(0,-1){.032895}}
\multiput(152.67,44.67)(-.032895,-.032895){20}{\line(-1,0){.032895}}
\multiput(151.35,43.35)(-.032895,-.032895){20}{\line(-1,0){.032895}}
\multiput(150.03,42.03)(-.032895,-.032895){20}{\line(0,-1){.032895}}
\multiput(148.72,40.72)(-.032895,-.032895){20}{\line(-1,0){.032895}}
\multiput(147.4,39.4)(-.032895,-.032895){20}{\line(-1,0){.032895}}
\multiput(146.09,38.09)(-.032895,-.032895){20}{\line(0,-1){.032895}}
\multiput(144.77,36.77)(-.032895,-.032895){20}{\line(-1,0){.032895}}
\multiput(143.46,35.46)(-.032895,-.032895){20}{\line(0,-1){.032895}}
\multiput(142.14,34.14)(-.032895,-.032895){20}{\line(0,-1){.032895}}
\multiput(140.82,32.82)(-.032895,-.032895){20}{\line(-1,0){.032895}}
\multiput(139.51,31.51)(-.032895,-.032895){20}{\line(-1,0){.032895}}
\multiput(138.19,30.19)(-.032895,-.032895){20}{\line(-1,0){.032895}}
\multiput(136.88,28.88)(-.032895,-.032895){20}{\line(-1,0){.032895}}
\multiput(135.56,27.56)(-.032895,-.032895){20}{\line(0,-1){.032895}}
\multiput(134.25,26.25)(-.032895,-.032895){20}{\line(-1,0){.032895}}
\put(23,10){\line(4,1){32}} \put(27,26){\line(4,1){32}}
\put(55,18){\line(1,4){8}} \put(23,10){\line(1,4){8}}
\put(31,42){\line(4,1){32}}
\multiput(22.93,9.93)(.033333,.033333){20}{\line(1,0){.033333}}
\multiput(24.26,11.26)(.033333,.033333){20}{\line(1,0){.033333}}
\multiput(25.6,12.6)(.033333,.033333){20}{\line(0,1){.033333}}
\multiput(26.93,13.93)(.033333,.033333){20}{\line(1,0){.033333}}
\multiput(28.26,15.26)(.033333,.033333){20}{\line(1,0){.033333}}
\multiput(29.6,16.6)(.033333,.033333){20}{\line(0,1){.033333}}
\multiput(30.93,17.93)(.033333,.033333){20}{\line(1,0){.033333}}
\multiput(32.26,19.26)(.033333,.033333){20}{\line(0,1){.033333}}
\multiput(33.6,20.6)(.033333,.033333){20}{\line(0,1){.033333}}
\multiput(34.93,21.93)(.033333,.033333){20}{\line(0,1){.033333}}
\multiput(36.26,23.26)(.033333,.033333){20}{\line(0,1){.033333}}
\multiput(37.6,24.6)(.033333,.033333){20}{\line(0,1){.033333}}
\multiput(38.93,25.93)(.033333,.033333){20}{\line(0,1){.033333}}
\multiput(40.26,27.26)(.033333,.033333){20}{\line(0,1){.033333}}
\multiput(41.6,28.6)(.033333,.033333){20}{\line(1,0){.033333}}
\multiput(42.93,29.93)(.033333,.033333){20}{\line(0,1){.033333}}
\multiput(44.26,31.26)(.033333,.033333){20}{\line(0,1){.033333}}
\multiput(45.6,32.6)(.033333,.033333){20}{\line(1,0){.033333}}
\multiput(46.93,33.93)(.033333,.033333){20}{\line(0,1){.033333}}
\multiput(48.26,35.26)(.033333,.033333){20}{\line(0,1){.033333}}
\multiput(49.6,36.6)(.033333,.033333){20}{\line(1,0){.033333}}
\multiput(50.93,37.93)(.033333,.033333){20}{\line(0,1){.033333}}
\multiput(52.26,39.26)(.033333,.033333){20}{\line(0,1){.033333}}
\multiput(53.6,40.6)(.033333,.033333){20}{\line(1,0){.033333}}
\multiput(54.93,41.93)(.033333,.033333){20}{\line(0,1){.033333}}
\multiput(56.26,43.26)(.033333,.033333){20}{\line(0,1){.033333}}
\multiput(57.6,44.6)(.033333,.033333){20}{\line(1,0){.033333}}
\multiput(58.93,45.93)(.033333,.033333){20}{\line(0,1){.033333}}
\multiput(60.26,47.26)(.033333,.033333){20}{\line(0,1){.033333}}
\multiput(61.6,48.6)(.033333,.033333){20}{\line(1,0){.033333}}
\multiput(30.93,41.93)(.032432,-.032432){20}{\line(1,0){.032432}}
\multiput(32.23,40.63)(.032432,-.032432){20}{\line(1,0){.032432}}
\multiput(33.52,39.34)(.032432,-.032432){20}{\line(1,0){.032432}}
\multiput(34.82,38.04)(.032432,-.032432){20}{\line(0,-1){.032432}}
\multiput(36.12,36.74)(.032432,-.032432){20}{\line(1,0){.032432}}
\multiput(37.42,35.44)(.032432,-.032432){20}{\line(0,-1){.032432}}
\multiput(38.71,34.15)(.032432,-.032432){20}{\line(0,-1){.032432}}
\multiput(40.01,32.85)(.032432,-.032432){20}{\line(0,-1){.032432}}
\multiput(41.31,31.55)(.032432,-.032432){20}{\line(1,0){.032432}}
\multiput(42.61,30.25)(.032432,-.032432){20}{\line(0,-1){.032432}}
\multiput(43.9,28.96)(.032432,-.032432){20}{\line(1,0){.032432}}
\multiput(45.2,27.66)(.032432,-.032432){20}{\line(1,0){.032432}}
\multiput(46.5,26.36)(.032432,-.032432){20}{\line(0,-1){.032432}}
\multiput(47.79,25.06)(.032432,-.032432){20}{\line(1,0){.032432}}
\multiput(49.09,23.77)(.032432,-.032432){20}{\line(0,-1){.032432}}
\multiput(50.39,22.47)(.032432,-.032432){20}{\line(1,0){.032432}}
\multiput(51.69,21.17)(.032432,-.032432){20}{\line(0,-1){.032432}}
\multiput(52.98,19.88)(.032432,-.032432){20}{\line(1,0){.032432}}
\multiput(54.28,18.58)(.032432,-.032432){20}{\line(0,-1){.032432}}
\put(22.5,10.5){\makebox(0,0)[rt]{A}}
\put(55,17){\makebox(0,0)[lt]{C}}
\put(63.5,50.5){\makebox(0,0)[lb]{D}}
\put(43.25,33.25){\makebox(0,0)[cc]{F}}
\put(30,42){\makebox(0,0)[rb]{B}}
\put(26,25){\makebox(0,0)[rc]{}}
\put(26,26){\makebox(0,0)[rc]{E}}
\put(42,2){\makebox(0,0)[cc]{(a)}}
\end{picture}

\caption{(a) Minkowski square, (b) similar triangles and (c)
causal domain.} \label{similar}
\end{figure}

We already remarked that an inertial particle trajectory is
represented by a timelike line segment. A second particle at rest
with respect to the first corresponds to a parallel line segment,
as illustrated by the two timelike sides AB and CD of the
parallelogram in Fig.~\ref{similar}a. The diagonals of this
parallelogram are light rays, so a light ray from A reaches point
F in the center, 
and is reflected back to B. Thus an observer
along AB (or CD) considers F to be simultaneous with the midpoint
E between A and B. Similarly all other points on the line EF are
considered simultaneous by this observer. Since EF is related to
AC by a translation, the points on AC are also simultaneous with
respect to AB. The segments AB and AC are said to be {\it
Minkowski-perpendicular}. A parallelogram like ABCD, with
lightlike diagonals, is a {\it Minkowski square}. The angle
between Minkowski-perpendicular lines is bisected by a light ray,
but this property holds only when the diagram is scaled so that
the light rays are perpendicular.

A pair of timelike lines that are not parallel represent inertial
particles with a relative velocity. The principle of relativity
demands that all such lines be equivalent. Fig.~\ref{similar}b
depicts two triangles OGH and OIJ, each consisting of two timelike
sides and one lightlike side. The directions of the timelike sides
are the same for the two triangles. In order that neither
direction be preferred the ratios of the corresponding proper
times must be equal, i.e. 
\beq
\frac{\rm (OH)_m}{\rm (OG)_m}=\frac{\rm (OJ)_m}{\rm (OI)_m}, 
\eeq
where for example ${\rm (OG)_m}$ denotes 
the proper time along OG.\footnote{This ratio is nothing but the
relativistic Doppler shift factor relating the reference frames
determined by the two timelike lines. Bondi's k-calculus
\cite{Bondi} is a presentation of special relativity in which this
ratio---the k-factor---is given the central role.} 
The two triangles are therefore
{\it similar} in Minkowski space.

Right triangles KML and KNL, each with two lightlike sides and a
timelike hypotenuse, are shown in Fig \ref{similar}c.  We call
these {\it null triangles}. Together they make up the rectangle
formed by the two pairs of light rays departing from the endpoints
K, L of the timelike segment. We call this rectangle the {\it
causal domain} of the timelike segment. It is also the causal
domain of the spacelike segment given by the other diagonal (MN)
of the rectangle.

\subsection*{Minkowski interval and Euclidean area}

For a vertical or horizontal segment it is easy to see that the
causal domain is a (Euclidean) square whose area is proportional
to the square (second power) of the proper time or length of the
segment, respectively.\footnote{If the horizontal and vertical
scales are chosen respectively to equal the Minkowski length and
the speed of light times the Minkowski time, then the
proportionality factor is one-half.} The same turns out to be true
for segments that are neither vertical nor horizontal, that is,
\begin{quote}
{\it The square of the proper time or length along any segment is
equal to the (Euclidean) area of its causal domain times a {\it
fixed} proportionality constant.}
\end{quote}
The same proportionality holds also for the area of the square
built on a timelike segment, since according to
Fig.~\ref{similar}a that area is always twice the area of the
causal domain. We now give a proof of this statement.

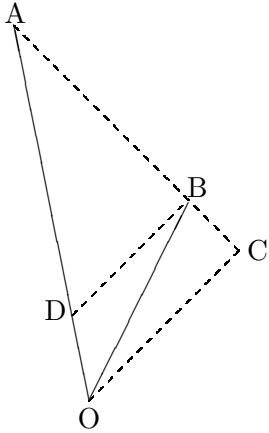
\begin{figure}
\unitlength 1mm 
\linethickness{0.4pt}
\ifx\plotpoint\undefined\newsavebox{\plotpoint}\fi 
\begin{picture}(81,61)(0,0)
\put(60,10){\line(-1,5){10}}
\put(60,10){\line(1,2){13.2}} \put(60,9){\makebox(0,0)[ct]{O}}
\put(57,22){\makebox(0,0)[rc]{D}}
\put(50.2,61.5){\makebox(0,0)[cc]{A}}
\put(73,37){\makebox(0,0)[lb]{B}}
\put(81,30){\makebox(0,0)[lc]{C}}
\multiput(79.93,29.93)(-.033333,.033333){20}{\line(-1,0){.033333}}
\multiput(78.6,31.26)(-.033333,.033333){20}{\line(-1,0){.033333}}
\multiput(77.26,32.6)(-.033333,.033333){20}{\line(0,1){.033333}}
\multiput(75.93,33.93)(-.033333,.033333){20}{\line(-1,0){.033333}}
\multiput(74.6,35.26)(-.033333,.033333){20}{\line(0,1){.033333}}
\multiput(73.26,36.6)(-.033333,.033333){20}{\line(0,1){.033333}}
\multiput(71.93,37.93)(-.033333,.033333){20}{\line(-1,0){.033333}}
\multiput(70.6,39.26)(-.033333,.033333){20}{\line(0,1){.033333}}
\multiput(69.26,40.6)(-.033333,.033333){20}{\line(0,1){.033333}}
\multiput(67.93,41.93)(-.033333,.033333){20}{\line(-1,0){.033333}}
\multiput(66.6,43.26)(-.033333,.033333){20}{\line(0,1){.033333}}
\multiput(65.26,44.6)(-.033333,.033333){20}{\line(-1,0){.033333}}
\multiput(63.93,45.93)(-.033333,.033333){20}{\line(-1,0){.033333}}
\multiput(62.6,47.26)(-.033333,.033333){20}{\line(0,1){.033333}}
\multiput(61.26,48.6)(-.033333,.033333){20}{\line(0,1){.033333}}
\multiput(59.93,49.93)(-.033333,.033333){20}{\line(-1,0){.033333}}
\multiput(58.6,51.26)(-.033333,.033333){20}{\line(0,1){.033333}}
\multiput(57.26,52.6)(-.033333,.033333){20}{\line(0,1){.033333}}
\multiput(55.93,53.93)(-.033333,.033333){20}{\line(-1,0){.033333}}
\multiput(54.6,55.26)(-.033333,.033333){20}{\line(0,1){.033333}}
\multiput(53.26,56.6)(-.033333,.033333){20}{\line(-1,0){.033333}}
\multiput(51.93,57.93)(-.033333,.033333){20}{\line(0,1){.033333}}
\multiput(50.6,59.26)(-.033333,.033333){20}{\line(0,1){.033333}}
\put(72.93,35.93){\line(0,1){0}}
\multiput(57.75,21.28)(.033684,.033684){19}{\line(0,1){.033684}}
\multiput(59.03,22.56)(.033684,.033684){19}{\line(0,1){.033684}}
\multiput(60.31,23.84)(.033684,.033684){19}{\line(1,0){.033684}}
\multiput(61.59,25.12)(.033684,.033684){19}{\line(0,1){.033684}}
\multiput(62.87,26.4)(.033684,.033684){19}{\line(0,1){.033684}}
\multiput(64.15,27.68)(.033684,.033684){19}{\line(1,0){.033684}}
\multiput(65.43,28.96)(.033684,.033684){19}{\line(0,1){.033684}}
\multiput(66.71,30.24)(.033684,.033684){19}{\line(0,1){.033684}}
\multiput(67.99,31.52)(.033684,.033684){19}{\line(0,1){.033684}}
\multiput(69.27,32.8)(.033684,.033684){19}{\line(0,1){.033684}}
\multiput(70.55,34.08)(.033684,.033684){19}{\line(0,1){.033684}}
\multiput(71.83,35.36)(.033684,.033684){19}{\line(0,1){.033684}}
\multiput(59.93,9.93)(.032258,.032258){20}{\line(1,0){.032258}}
\multiput(61.22,11.22)(.032258,.032258){20}{\line(1,0){.032258}}
\multiput(62.51,12.51)(.032258,.032258){20}{\line(1,0){.032258}}
\multiput(63.8,13.8)(.032258,.032258){20}{\line(0,1){.032258}}
\multiput(65.09,15.09)(.032258,.032258){20}{\line(1,0){.032258}}
\multiput(66.38,16.38)(.032258,.032258){20}{\line(0,1){.032258}}
\multiput(67.67,17.67)(.032258,.032258){20}{\line(0,1){.032258}}
\multiput(68.96,18.96)(.032258,.032258){20}{\line(0,1){.032258}}
\multiput(70.25,20.25)(.032258,.032258){20}{\line(0,1){.032258}}
\multiput(71.54,21.54)(.032258,.032258){20}{\line(0,1){.032258}}
\multiput(72.83,22.83)(.032258,.032258){20}{\line(0,1){.032258}}
\multiput(74.12,24.12)(.032258,.032258){20}{\line(1,0){.032258}}
\multiput(75.41,25.41)(.032258,.032258){20}{\line(0,1){.032258}}
\multiput(76.7,26.7)(.032258,.032258){20}{\line(0,1){.032258}}
\multiput(77.99,27.99)(.032258,.032258){20}{\line(1,0){.032258}}
\multiput(79.28,29.28)(.032258,.032258){20}{\line(0,1){.032258}}
\end{picture}
\vskip-5\unitlength
\caption{The area of null triangles is
proportional to the square of their timelike sides.} \label{jwyo}
\end{figure}

Figure \ref{jwyo} shows two timelike segments of different
directions (velocities), OA and OB, with A and B lightlike
related. We construct the null triangles OCB and OCA that
represent the right half of the two segments' causal domains, and
the null line DB.

Let $\cal A$(OA) be the area of OA's causal domain, which is twice
the area of the null triangle OCA, and similarly for $\cal A$(OB).
Because the two null triangles have the common base OC, and
because OC and DB are parallel, we have the proportionality,
\beq{\rm {{\cal A}(OA)\over {\cal A}(OB)} = {(OA)_e\over (OD)_e}=
{(OA)_m\over (OD)_m} ,}
\eeq
where the subscripts e and m refer to Euclidean length and
Minkowskian time respectively. Because the triangles ODB and OBA
are Minkowski-similar, we have
\beq
{\rm {(OA)_m\over (OB)_m} = {(OB)_m\over (OD)_m},
\quad {\rm hence} \quad {{\cal A}(OA)\over {\cal A}(OB)}
= {(OA)_m^2\over (OB)_m^2}.}
\eeq
In other words, the area ratio
of the two causal domains is equal to the ratio of the square of
the proper times along their timelike diagonals.

Our argument applies only for segments related by a {\it null}
line BA.
However, since Euclidean lengths along a single timelike segment
are proportional to the corresponding proper times, and the area
scales with the square of the Euclidean length, the result is
valid for any pair of causal domains.

\subsection*{Time dilation}

\begin{figure}[h]
\centering
\includegraphics[width=\textwidth]{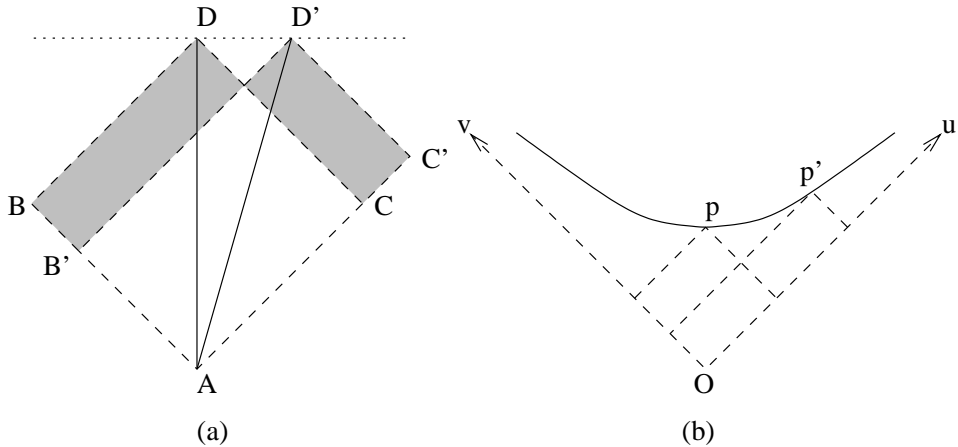}
\caption{(a) Time dilation (b) locus of points at fixed future timelike
interval from O.} 
\label{timedil}
\end{figure}

As a first application of the relation between area and spacetime
interval one can see immediately in Fig.~\ref{timedil}a that of
two timelike intervals AC, AC$'$ with the same vertical
projection, the slanted one has the shorter proper time. This is
the relativistic time dilation or twin effect. Fig.~\ref{timedil}a
shows half of the twin's round trip (a round trip would be
obtained by reflecting the figure about the dotted horizontal line):
the area of the rectangle AB$'$C$'$D$'$ is less than that of the
square ABCD, for the shaded areas have equal narrow width, but the
one that is part of the rectangle is shorter than the one that is
part of the square. As the relative velocity of the tilted segment
increases, its causal domain area goes to zero, and hence so does
its proper time.

The locus of spacetime points P, P$'$,\dots  that are at a
constant future timelike interval from a given origin O is shown
in Fig.~\ref{timedil}b. This locus is given by 
(OP)$^2\propto{\cal A}$(OP)$=uv$, 
where $u,\,v$ are the null coordinates of P, so the
locus is a {\em hyperbola}. Similarly the points at constant past
timelike separations, and at constant spacelike separations 
from O are also hyperbolae.

\subsection*{Spacetime Pythagoras theorem}

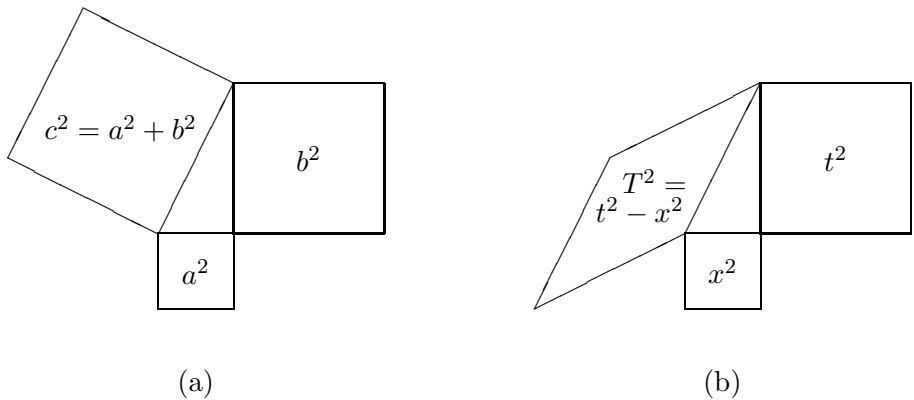
\begin{figure}
\unitlength 1mm \linethickness{0.4pt}
\begin{picture}(130.00,50.00)(10,0)
\put(30.05,10.00){\framebox(9.90,9.95)[cc]{$a^2$}}
\put(40.07,20.05){\framebox(19.9,19.9)[cc]{$b^2$}}
\put(25.00,34.00){\makebox(0,0)[cc]{$c^2=a^2+b^2$}}
\put(100.07,10.00){\framebox(9.95,10.00)[cc]{$x^2$}}
\put(110.10,20.05){\framebox(19.9,19.9)[cc]{$t^2$}}
\put(96.00,27.00){\makebox(0,0)[cc]{$T^2=$}}
\put(94.00,23.00){\makebox(0,0)[cc]{$t^2-x^2$}}
\put(35.00,0.00){\makebox(0,0)[cc]{(a)}}
\put(105.00,0.00){\makebox(0,0)[cc]{(b)}}
\put(30.00,20.00){\line(1,2){10.00}}
\put(40.00,40.00){\line(-2,1){20.00}}
\put(20.00,50.00){\line(-1,-2){10.00}}
\put(10.00,30.00){\line(2,-1){20.00}}
\put(100.00,20.00){\line(1,2){10.00}}
\put(110.00,40.00){\line(-2,-1){20.00}}
\put(90.00,30.00){\line(-1,-2){10.00}}
\put(80.00,10.00){\line(2,1){20.00}}
\end{picture}

\caption{Pythagorean theorem, (a) Euclidean (b) Minkowskian.}
\label{squares}
\end{figure}

Since spacetime intervals are determined by Euclidean areas in a
spacetime diagram, we can use Euclidean geometry to establish the
{\it spacetime Pythagoras theorem}. This is the
fundamental Minkowskian formula, relating time and space
measurements $t$ and $x$ of an interval by one observer to the
proper time measurement $T$ of that interval by
another observer,
\begin{equation}
T^2=t^2-x^2.
\label{sP}
\end{equation}
Figure \ref{squares} shows a geometrical view of the terms in this
equation and compares it to the familiar Euclidean interpretation
of the Pythagorean theorem
for right triangles. In the Minkowski case, two sides
of the triangle are Minkowski-perpendicular, as appropriate
for the time and space components of the 
hypotenuse displacement defined by a given observer. 
The square on
the hypotenuse also has Minkowski-perpendicular sides, and is a
parallelogram with lightlike diagonals as explained above.
The example of Fig.~\ref{squares}b is a special case since the
triangle sides are also perpendicular in the
Euclidean sense. However, the principle of relativity implies
that if the theorem holds for this case it must hold for any 
Minkowski right triangle.

There are many ways to prove the spacetime Pythagoras theorem,
just as there are in the Euclidean case \cite{book}. Here we
mention just two, the first using causal domains and the second
using the more traditional squares on the sides of the triangle.
Another proof, using spacetime tiling, was given in
Ref.~\cite{DieterSamos}.

\begin{figure}[t]
\centering
\includegraphics{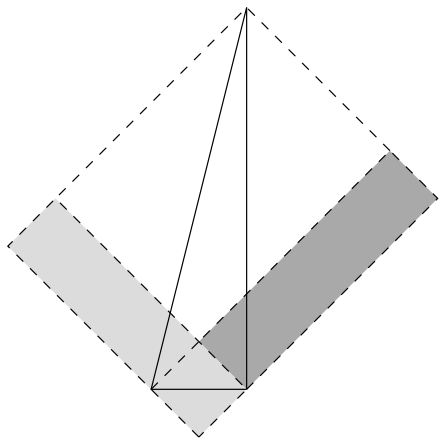}

\caption{A spacetime triangle and the causal domains of its
sides.}
\label{proof1}
\end{figure}

The first proof is illustrated in Fig.~\ref{proof1}, which shows a
right triangle with a vertical timelike side, a horizontal
spacelike base, and a timelike hypotenuse, together with
their causal domains. The intersection of the domain of the 
the hypotenuse with that of the vertical timelike side is the 
white rectangle. The dark grey rectangle is the rest of the 
domain of the vertical side. The light grey rectangle is the union of
the rest of the domain of the hypotenuse and the domain of the
spacelike side. The light and dark grey rectangles have the same
length and width, and therefore the same area, 
since the causal domains of the vertical
and horizontal sides of the triangle are Euclidean squares.
Hence the domain area on the vertical timelike side is
equal to the sum of those on the hypotenuse and spacelike side.
Since the areas are proportional to the squared Minkowski lengths,
this establishes (\ref{sP}).

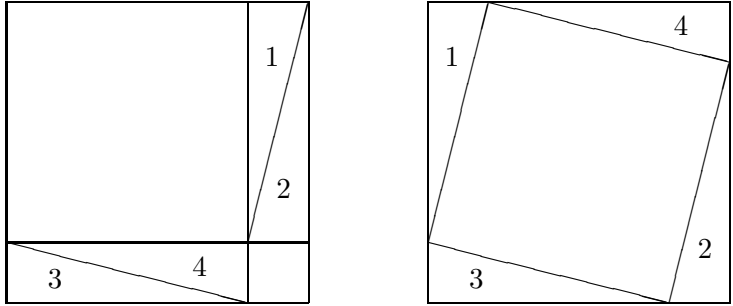
\begin{figure}[h]
\unitlength 0.80mm
\linethickness{0.4pt}

\begin{picture}(130.00,60.00)(-5,0)
\put(10.00,20.00){\line(1,0){50.00}}
\put(50.00,10.00){\line(0,1){50.00}}
\put(10.00,10.00){\framebox(50.00,50.00)[cc]{}}
\put(50.00,20.00){\line(1,4){10.00}}
\put(80.00,10.00){\framebox(50.00,50.00)[cc]{}}
\put(54.00,51.00){\makebox(0,0)[cc]{1}}
\put(56.00,29.00){\makebox(0,0)[cc]{2}}
\put(10.00,20.00){\line(4,-1){40.00}}
\put(80.00,20.00){\line(1,4){10.00}}
\put(84.00,51.00){\makebox(0,0)[cc]{1}}
\put(120.00,10.00){\line(1,4){10.00}}
\put(126.00,19.00){\makebox(0,0)[cc]{2}}
\put(18.00,14.00){\makebox(0,0)[cc]{3}}
\put(42.00,16.00){\makebox(0,0)[cc]{4}}
\put(80.00,20.00){\line(4,-1){40.00}}
\put(88.00,14.00){\makebox(0,0)[cc]{3}}
\put(90.00,60.00){\line(4,-1){40.00}}
\put(122.00,56.00){\makebox(0,0)[cc]{4}}
\end{picture}
\vskip-10\unitlength
\caption{Proof by rearrangement in the
Euclidean case.} \label{proof2a}
\end{figure}

The second proof uses squares on the sides of the triangle, and is
more closely analogous to a Euclidean proof. The latter is shown
in Fig.~\ref{proof2a}. Rearrangement of the four triangles as
shown converts the empty area from the two squares on the smaller
sides to the single square on the hypotenuse. This is perhaps the
most elegant and elementary geometric proof of the Euclidean
Pythagoras theorem.

\begin{figure}
\unitlength 0.80mm
\linethickness{0.4pt}
\begin{picture}(130.00,60.00)(-5,0)
\put(10.00,20.00){\line(1,0){50.00}}
\put(50.00,10.00){\line(0,1){50.00}}
\put(10.00,10.00){\framebox(50.00,50.00)[cc]{}}
\put(50.00,20.00){\line(1,4){10.00}}
\put(10.00,10.00){\line(4,1){40.00}}
\put(80.00,10.00){\framebox(50.00,50.00)[cc]{}}
\put(80.00,50.00){\line(1,0){10.00}}
\put(90.00,50.00){\line(0,1){10.00}}
\put(120.00,10.00){\line(0,1){10.00}}
\put(120.00,20.00){\line(1,0){10.00}}
\put(80.00,10.00){\line(1,4){10.00}}
\put(90.00,50.00){\line(4,1){40.00}}
\put(130.00,60.00){\line(-1,-4){10.00}}
\put(120.00,20.00){\line(-4,-1){40.00}}
\put(54.00,51.00){\makebox(0,0)[cc]{1}}
\put(84.00,41.00){\makebox(0,0)[cc]{1}}
\put(126.00,29.00){\makebox(0,0)[cc]{2}}
\put(56.00,29.00){\makebox(0,0)[cc]{2}}
\put(18.00,16.00){\makebox(0,0)[cc]{3}}
\put(98.00,56.00){\makebox(0,0)[cc]{3}}
\put(41.00,14.00){\makebox(0,0)[cc]{4}}
\put(111.00,14.00){\makebox(0,0)[cc]{4}}
\put(50.00,10.00){\rule{10.00\unitlength}{10.00\unitlength}}
\put(120.00,10.00){\rule{10.00\unitlength}{10.00\unitlength}}
\end{picture}
\vskip-10\unitlength
\caption{Proof by rearrangement in the
Minkowskian case.} \label{proof2b}
\end{figure}
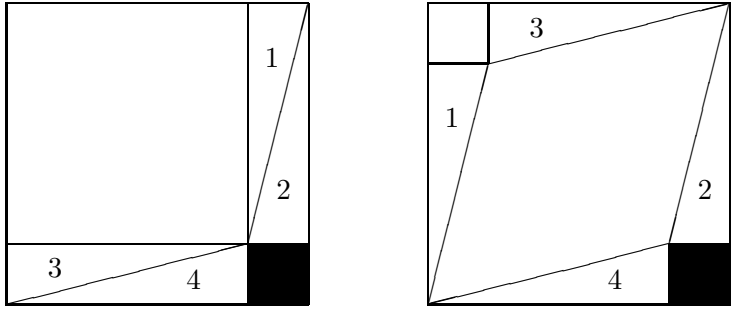

The corresponding proof in the Minkowski case is shown in
Fig.~\ref{proof2b}. The figure on the left differs from that in
Fig.~\ref{proof2a} only by the orientation of triangles 3 and
4.\footnote{We flipped these orientations so that the
rearrangement in Fig.~\ref{proof2b} involves just sliding, with no
flipping.} Rearrangement of the four triangles as shown converts
the empty unshaded area from the larger square to the rhombus plus
the smaller square. Thus the rhombus area is the {\it difference}
of the areas of the larger and smaller squares. The rhombus is
also the Minkowskian square on the hypotenuse, since it is a
parallelogram with lightlike diagonals. Therefore we have again
established (\ref{sP}).

By stretching Figs.~\ref{proof1} and \ref{proof2b}
in one null direction and shrinking by an equal factor in the other
null direction ($u$ resp.~$v$-directions of Fig.~\ref{timedil}) we
preserve all null directions and hence Minkowski-perpendicularity
and squares, as well as areas; and in this way we can obtain a
general 
Minkowski right
triangle, to which the spacetime
Pythagoras theorem applies.\footnote{In fact, the transformation
described is just a Lorentz transformation.} The so-transformed
Figure \ref{proof1} does not lend itself directly to the proof we
gave above because the shaded rectangles are no longer congruent
(in the Euclidean sense), although they do have equal area.
However, the second proof does still work for the transformed
Fig.~\ref{proof2b}, as shown in Fig.~{\ref{prooftrans}, because
all four transformed triangles are congruent.

\begin{figure}[h]
\unitlength 0.90mm
\linethickness{0.4pt}
\begin{picture}(125.00,50.00)(-5,0)
\put(60.00,0.00){\line(-4,1){40.00}}
\put(60.00,0.00){\line(-1,4){10.00}}
\put(10.00,50.00){\line(1,-4){10.00}}
\put(10.00,50.00){\line(4,-1){40.00}}
\put(20.00,10.00){\line(6,1){25.83}}
\put(45.83,14.31){\line(1,6){4.24}}
\put(57.12,11.36){\line(-4,1){40.00}}
\put(48.64,2.88){\line(-1,4){10.00}}
\thicklines
\put(45.91,13.94){\line(4,-1){11.36}}
\put(57.00,11.36){\line(1,-4){2.80}}
\put(59.92,0.15){\line(-4,1){10.98}}
\put(48.90,2.90){\line(-1,4){2.76}}
\thinlines
\put(46.93,13.20){\circle*{1.20}}
\put(48.93,2.93){\rule{8.00\unitlength}{8.40\unitlength}}
\put(48.87,11.23){\rule{4.00\unitlength}{1.03\unitlength}}
\put(48.90,12.17){\rule{2.03\unitlength}{0.60\unitlength}}
\put(52.80,11.30){\rule{2.07\unitlength}{0.57\unitlength}}
\put(49.00,11.00){\circle*{4.56}}
\put(47.53,12.57){\circle*{2.20}}
\put(51.30,11.17){\circle*{3.04}}
\put(52.90,10.93){\circle*{2.71}}
\put(55.07,10.40){\circle*{2.63}}
\put(47.73,7.13){\rule{1.03\unitlength}{4.00\unitlength}}
\put(47.23,9.07){\rule{0.60\unitlength}{2.03\unitlength}}
\put(48.13,5.13){\rule{0.57\unitlength}{2.07\unitlength}}
\put(48.83,8.70){\circle*{3.04}}
\put(49.07,7.10){\circle*{2.71}}
\put(49.60,4.93){\circle*{2.63}}
\put(46.37,13.63){\circle*{0.39}}
\put(59.20,0.93){\circle*{1.20}}
\put(57.00,3.00){\circle*{4.56}}
\put(58.57,1.53){\circle*{2.20}}
\put(57.17,5.30){\circle*{3.04}}
\put(56.93,6.90){\circle*{2.71}}
\put(56.40,9.07){\circle*{2.63}}
\put(54.70,2.83){\circle*{3.04}}
\put(53.10,3.07){\circle*{2.71}}
\put(50.93,3.60){\circle*{2.63}}
\put(56.50,10.40){\circle*{1.47}}
\put(57.23,8.17){\circle*{1.20}}
\put(49.33,3.63){\circle*{1.40}}
\put(51.97,2.63){\circle*{1.07}}
\put(59.63,0.37){\circle*{0.41}}
\put(125.00,0.00){\line(-4,1){40.00}}
\put(125.00,0.00){\line(-1,4){10.00}}
\put(75.00,50.00){\line(1,-4){10.00}}
\put(75.00,50.00){\line(4,-1){40.00}}
\put(85.00,10.00){\line(6,1){25.83}}
\put(110.83,14.31){\line(1,6){4.24}}
\thicklines
\put(110.91,13.94){\line(4,-1){11.36}}
\put(122.00,11.36){\line(1,-4){2.80}}
\put(124.92,0.15){\line(-4,1){10.98}}
\put(113.80,2.90){\line(-1,4){2.76}}
\thinlines
\put(111.93,13.20){\circle*{1.20}}
\put(113.93,2.93){\rule{8.00\unitlength}{8.40\unitlength}}
\put(113.87,11.23){\rule{4.00\unitlength}{1.03\unitlength}}
\put(113.90,12.17){\rule{2.03\unitlength}{0.60\unitlength}}
\put(117.80,11.30){\rule{2.07\unitlength}{0.57\unitlength}}
\put(114.00,11.00){\circle*{4.56}}
\put(112.53,12.57){\circle*{2.20}}
\put(116.30,11.17){\circle*{3.04}}
\put(117.90,10.93){\circle*{2.71}}
\put(120.07,10.40){\circle*{2.63}}
\put(112.73,7.13){\rule{1.03\unitlength}{4.00\unitlength}}
\put(112.23,9.07){\rule{0.60\unitlength}{2.03\unitlength}}
\put(113.13,5.13){\rule{0.57\unitlength}{2.07\unitlength}}
\put(113.83,8.70){\circle*{3.04}}
\put(114.07,7.10){\circle*{2.71}}
\put(114.60,4.93){\circle*{2.63}}
\put(111.37,13.63){\circle*{0.39}}
\put(124.20,0.93){\circle*{1.20}}
\put(122.00,3.00){\circle*{4.56}}
\put(123.57,1.53){\circle*{2.20}}
\put(122.17,5.30){\circle*{3.04}}
\put(121.93,6.90){\circle*{2.71}}
\put(121.40,9.07){\circle*{2.63}}
\put(119.70,2.83){\circle*{3.04}}
\put(118.10,3.07){\circle*{2.71}}
\put(115.93,3.60){\circle*{2.63}}
\put(121.50,10.40){\circle*{1.47}}
\put(122.23,8.17){\circle*{1.20}}
\put(114.33,3.63){\circle*{1.40}}
\put(116.97,2.63){\circle*{1.07}}
\put(124.63,0.37){\circle*{0.41}}
\put(89.31,35.83){\line(6,1){25.44}}
\put(86.33,47.11){\line(1,-4){2.83}}
\put(89.17,35.78){\line(-4,1){11.28}}
\put(85.00,10.00){\line(1,6){4.31}}
\put(122.11,11.33){\line(-4,1){11.33}}
\put(110.78,14.17){\line(1,-4){2.82}}
\put(25.00,15.00){\makebox(0,0)[cb]{3}}
\put(94.00,41.00){\makebox(0,0)[cc]{3}}
\put(41.00,9.00){\makebox(0,0)[cc]{4}}
\put(106.00,9.00){\makebox(0,0)[cc]{4}}
\put(51.00,19.00){\makebox(0,0)[cc]{2}}
\put(116.00,19.00){\makebox(0,0)[cc]{2}}
\put(45.00,35.00){\makebox(0,0)[cc]{1}}
\put(84.00,31.00){\makebox(0,0)[cc]{1}}
\end{picture}
\caption{Rearrangement proof for a general
triangle.}
\label{prooftrans}
\end{figure}
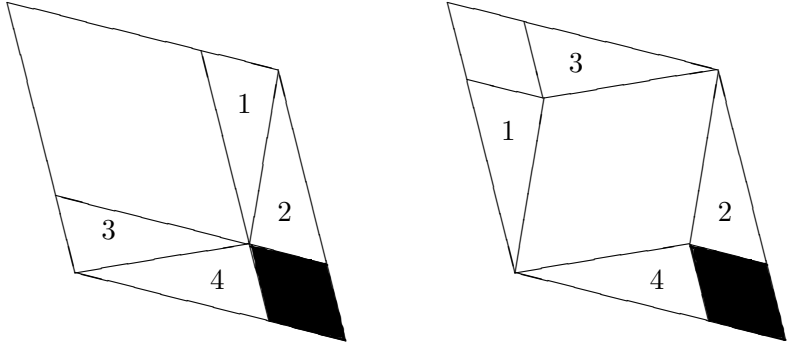

\subsection*{Minkowski area and higher dimensions}

Throughout this paper we have spoken of area only in the Euclidean
sense. Using the notion of Minkowski area,
we can give a more abstract proof of the proportionality
of Euclidean area and squared proper time, which also
generalizes to higher dimensional spacetimes, as follows.

In any dimension one can map Minkowski to Euclidean space
in a manner that preserves the translation symmetry 
and hence the straight lines, and 
Euclidean and Minkowskian length are proportional on a given line.
Moreover, since a translationally invariant volume element is
determined up to a constant scalar multiple, the 
image of any such Minkowskian volume element under any such
map is necessarily proportional to the Euclidean
volume element.

Now consider a timelike segment AB of proper time $T$
in an $n$-dimensional Minkowski space, and define
the causal domain of AB as 
the intersection of the future of A with the past of B.
The ratio of $T^n$ to the Minkowski volume of the 
corresponding casual domain is a dimensionless number. 
Since a Minkowski volume element does not determine any 
preferred timelike direction, this ratio must be the same for all
timelike intervals. The ratio of $T^n$ to the Euclidean
volume is thus also constant. This can be used to interpret
three-dimensional spacetime diagrams in much the same
way as we did here with two-dimensional ones.

\subsection*{Acknowledgements}
We would like to thank the  organizers of the Second Samos Meeting
on Cosmology, Geometry and Relativity, held at Pythagoreon, Samos, Greece
in September 1998, which stimulated this project. 
This work was supported in part by
the NSF under grants PHY-9800967 and PHY-0300710 at the University
of Maryland, and by the CNRS at the Institut
d'Astrophysique de Paris.

\subsection*{Note Added}
After this article was completed we became aware of some related work.
A simple geometric proof that the squared interval is proportional
to the area of the corresponding causal domain (``light rectangle") 
was given by Mermin~\cite{Mermin2}. His construction assumes the two intervals
being compared have the same length, and is hence more symmetric than
the one given in our article. A geometric proof of what we 
called the spacetime Pythagoras theorem was given as early as 1913 in 
Propositions XI and XXI of Ref.~\cite{WilsonLewis}. 
A proof by Liebscher presented with animated graphics  is 
available online~\cite{Liebscher1} (see also~\cite{Liebscher3}).  
We are grateful to R.~Salgado and D.~Liebscher for steering us to 
these references.

\end{document}